\documentclass[aps,prl,twocolumn,groupedaddress]{revtex4}
\usepackage{graphicx}
\begin{document}
\newcommand{\ud}{{\mathrm d}}
\newcommand{\sech}{\mathrm{sech}}

\title{Two-state theory of nonlinear Stochastic Resonance}


\author{Jes\'us Casado-Pascual, Jos\'e  G\'omez-Ord\'o\~nez,
Manuel Morillo}
\affiliation{F\'{\i}sica Te\'orica,
Universidad de Sevilla, Apartado de Correos 1065, Sevilla 41080, Spain}
\author{Peter H\"anggi}
\affiliation{Institut f\"ur Physik,
Universit\"at Augsburg - Universit\"atsstra\ss e 1, D-86135 Augsburg,
Germany}

\date{\today}

\begin{abstract}
An amenable, analytical two-state description of the {\em nonlinear}
population dynamics of a noisy bistable system driven by a rectangular
subthreshold signal is put forward. Explicit expressions for the driven
population dynamics, the correlation function (its coherent and incoherent
part), the signal-to-noise ratio (SNR) and the Stochastic Resonance (SR)
gain are obtained. Within a suitably chosen range of parameter values this
reduced description yields anomalous SR-gains exceeding unity and,
simultaneously, gives rise to a non-monotonic behavior of the SNR {\it
vs.} the noise strength. The analytical results agree well with those
obtained from numerical solutions of the Langevin equation.
\end{abstract}


\maketitle

The phenomenon of Stochastic Resonance (SR) attracts ever growing
interest due to its multi-facetted relevance for a variety of
noise-induced features in physics, chemistry, and the life sciences
\cite{PT96,RMP,Wiesenfeld98,Anish99,Chemphyschem02}. Several
SR-quantifiers have been used to characterize the response of a noisy
system to the action of time-periodic external forces. In particular,
the non-monotonic behavior of the output signal-to-noise ratio (SNR)
with the strength of the noise has been used widely. A dimensionless
quantity that measures the ``quality'' of the response with respect to
the input signal is the SR-gain defined as the ratio of the output SNR
over the input SNR. Ideally, one would wish to obtain the characteristic
amplification of the SR phenomenon \cite{RMP,PRA91} and, simultaneously,
SR-gains larger than unity. For superthreshold sinusoidal input signals
SR-gains larger than unity have been reported before \cite{haninc00}. In
recent analog \cite{Kish,ginmak00,ginmak01} and numerical
\cite{casgom03,casgom03new} simulations of noisy bistable systems driven
by {\it subthreshold} multifrequency input forces, surprisingly large
SR-gains larger than unity have been established.

In order to clarify the conditions under which these anomalous large
SR-gains occur, it would be interesting to propose simplified models,
amenable to analytical treatment, which describe this rich behavior of
the response.  A detailed proof that SR-gains larger than unity are
incompatible with Linear Response Theory (LRT) has been presented in
\cite{casgom03}. Thus, any theoretical explanation of the simultaneous
existence of SR and anomalous large gains is rooted in the response
beyond LRT.

The main focus of this work is to present such a simplified two-state
description of the nonlinear dynamics of a noisy, symmetric bistable
system driven by a rectangular subthreshold signal. A two-state
description of SR has been considered previously in the pioneering work by
McNamara and Wiesenfield for sinusoidal input signals \cite{McNWie89}. In
clear contrast to their work, however, we will here {\em not} linearize
the transition probabilities in the strength of the applied force.  In
doing so, we put forward explicit analytical expressions for the driven
population probabilities, the nonlinear correlation function (its coherent
and incoherent part), the SNR and the SR-gain.  These novel nonlinear
response results come forth solely  because the rectangular signal -- in
contrast to sinusoidal driving -- involves two force values only.

To start, let us consider a system characterized by a single degree of
freedom, $x$, whose dynamics (in dimensionless units)  is governed by the
Langevin equation
\begin{equation}
\label{langev} \dot{x}(t)=-U'\left[ x(t),t \right]+\xi(t),
\end{equation}
where $\xi (t)$ is Gaussian white noise of zero mean with $\langle
\xi(t)\xi(s)\rangle = 2D\delta(t-s)$, and $ -U'(x,t)$ represents the
force stemming from the time-dependent, archetype bistable potential
\begin{equation}
\label{potential}
U(x,t)=\frac{x^4}{4}-\frac{x^2}{2}-F(t)x.
\end{equation}
We will restrict our study to a periodic rectangular driving force with
period $T$,
\begin{equation}
F(t)=(-1)^{n(t)} A,
\end{equation}
where $n(t)=\lfloor 2\, t/T \rfloor$, $\lfloor z \rfloor$ being the
floor function of $z$, i.e., the greatest integer less than or equal to
$z$. In other words, $F(t)=A$ [$F(t)=-A$] if $t\in [n T/2, (n+1)T/2)$
with $n$ even (odd). Our focus is on subthreshold signals; more
precisely, we will assume that $A<A_{th}=\sqrt{4/27}$ where $A_{th}$ is
the static threshold value (the dynamical threshold value always exceeds
this adiabatic threshold $A_{th}$). In this case, the potential $U(x,t)$
presents two minima at $x_1(t)<0$ and $x_2(t)>0$, and a maximum at
$x_M(t)$. Because the potential fulfills the symmetry property
$U(x,t+T/2)=U(-x,t)$, then $x_M(t)=(-1)^{n(t)}x_M(0)$, and
\begin{equation}
\label{minima}
x_j(t)=(-1)^j \frac{\Delta x(0)}{2}-(-1)^{n(t)}\frac{x_M(0)}{2},
\end{equation}
where $\Delta x(0)=x_2(0)-x_1(0)$. Additionally, we have taken into
account the symmetry relation $x_1(t)+x_2(t)+x_M(t)=0$.

Throughout the following, we will assume that the noise strength, $D$,
is sufficiently small so that the intrawell relaxation time scale is
negligible compared with the time scale associated to the interwell
transitions and, as well, the driving time scale $T/2$. In this case,
and for sufficiently slow driving \cite {Astumian99,Talkner03}, the
exact Langevin dynamics can be approximated by a nonstationary,
Markovian two-state description of the form
\begin{eqnarray}
\label{ME1}
\dot{P}(1,t)&=&-\gamma_1(t)P(1,t)+\gamma_2
(t)P(2,t), \\
\label{ME2}
\dot{P}(2,t)&=&-\gamma_2(t)P(2,t)+\gamma_1
(t)P(1,t).
\end{eqnarray}
Here, $P(1,t)$ and $P(2,t)$ denote, respectively, the populations to the
left and to the right of the maximum position, $x_M(t)$, and
$\gamma_j(t)$ is the Kramers' rate of escape \cite {RMP90} from the well
$j$ at time $t$, i.e.,
\begin{equation}
\label{Kramers}
\label{krates}
\gamma_j(t)=\frac{\omega_j(t) \, \omega_M(t)}{2 \pi}
\exp\left\{-\frac{U\left[x_M(t),t\right]-U\left[x_j(t),t\right]}{D}
\right\},
\end{equation}
where $\omega_j(t)=\sqrt{U''[x_j(t),t]}=\sqrt{3 [x_j(t)]^2-1}$ and
$\omega_M(t)=\sqrt{|U''[x_M(t),t]|}=\sqrt{1-3
[x_M(t)]^2}=\omega_M(0)$. According to the above mentioned symmetry
property of the potential, $\gamma_j(t)$ can be expressed in the form
\begin{equation}
\gamma_j(t)=\frac{\Gamma}{2} \left[1-(-1)^{n(t)+j} \Delta P_{eq}(0) \right],
\end{equation}
where $\Gamma=\gamma_1(0)+\gamma_2(0)=\gamma_1(t)+\gamma_2(t)$, and
$\Delta P_{eq}(0)=P_{eq}(2,0)-P_{eq}(1,0)$, $P_{eq}(j,0)$ being the
equilibrium population of the state $j$ corresponding to the rates taken
at time $t=0$, i.e., $P_{eq}(j,0)=\left[\delta_{j,1}\,
\gamma_2(0)+\delta_{j,2} \,\gamma_1(0)\right]/\Gamma$. Eqs.~(\ref{ME1})
and (\ref{ME2}) describe the evolution of the population between two
consecutive changes of shape of the potential. These populations can be
discontinuous at $t=n T/2$, as a result of the sudden change in the
location of the maximum at those instants of time. Therefore, in order
to complete the description, jump conditions for the populations at $t=n
T/2$ have to be added. For our situation with a rectangular signal we
shall assume that the probability distribution before a change of the
potential is sufficiently well localized around the minima that there is
almost no probability transfer from one well to the other as the maximum
location changes. In this case, the populations can be considered to be
continuous at $t=n T/2$, i.e.,
\begin{equation}
\label{jumpcond}
\lim_{t\rightarrow \frac{n T}{2}-} P(j,t)=\lim_{t\rightarrow \frac{n
T}{2}+} P(j,t)= P\left(j,n T/2\right).
\end{equation}
This relation becomes exact at asymptotic weak noise. At finite weak noise
strength $D$  those finite jump conditions induce for the correlation
function some small, weakly nonanalytic structures which in turn may cause
dip-like features around even numbered multiples of the angular driving
frequency $\Omega$. Indeed, such dips in the spectrum are well known to
occur at weak noise for time-continuous, e.g. sinusoidally rocked bistable
systems in the nonlinear response regime \cite {1994}; note, however, that
for our case with  a constant force (or potential) the source of the
nonanalytic structure is of different origin. For rectangular driving
forces and  noise strength values leading to SR-gains larger than unity,
the incoherent part of the correlation decays very rapidly on the time
scale $T/2$ so that these small corrections stemming from the
approximation in (\ref{jumpcond}) can safely be neglected. After using the
normalization condition $P_1(1,t)+P_1(2,t)=1$, Eqs.~(\ref{ME1}) and
(\ref{ME2}) yield
\begin{equation}
\label{MEreduced}
\dot{P}(1,t)=-\Gamma \, P(1,t)+\gamma_{2}(t).
\end{equation}
Upon observing that $\gamma_{2}(t)$ remains constant between two
consecutive changes of the potential, the solution of
Eq.~(\ref{MEreduced}) can be expressed as
\begin{equation}
\label{gensol}
P(1,t)=\frac{\gamma_{2}(t)}{\Gamma}+\left\{P\left[1,n(t) T/2\right]
-\frac{\gamma_{2}(t)}{\Gamma}\right\}e^{-\Gamma
\left[t-\frac{n(t)T}{2}\right]} .
\end{equation}

With this relation, it is straightforward to evaluate the time-periodic,
asymptotic long-time solution of Eq.~(\ref{MEreduced}), i.e.
$P_{\infty}(1,t)$. In order to do so, we calculate the values
$P_{\infty}(1,n T/2)$ by making use of the symmetry property
$P_{\infty}(1,t\pm T/2)=P_{\infty}(2,t)=1-P_{\infty}(1,t)$, as well as
of Eq.~(\ref{jumpcond}). After inserting the result in
Eq.~(\ref{gensol}), one finds the $T$-periodic solution
\begin{eqnarray}
\label{persol}
P_{\infty}(1,t)&=&\frac{1}{2} \left[1-(-1)^{n(t)} \Delta
P_{eq}(0)\right]\nonumber\\&& +(-1)^{n(t)} \Delta P_{eq}(0)\,
\frac{e^{-\Gamma \left[t-\frac{n(t)T}{2}\right]}}{1+e^{-\frac{\Gamma
T}{2}}}.
\end{eqnarray}
The average of the coordinate in the long-time limit, $\langle x(t)
\rangle_{\infty}^{(TS)}=\sum_{j=1}^2 x_j(t)P_{\infty}(j,t)$, is
evaluated to read
\begin{eqnarray}
\label{TSaverag}
\langle x(t) \rangle_{\infty}^{(TS)}&=& (-1)^{n(t)} \Bigg\{\langle x(0)
\rangle_{eq} -\left[x_M(0)+2\langle x(0)\rangle_{eq}
\right]\nonumber \\
&&\times\,\frac{e^{-\Gamma \left[t-\frac{n(t) T}{2} \right]}}{1+
e^{-\frac{\Gamma T}{2}}} \Bigg\},
\end{eqnarray}
where $\langle x(0)\rangle_{eq}=\sum_{j=1}^2x_j(0)P_{eq}(j,0)$.

The conditional probability $P(1,t|j,t_0)$ can also be calculated using
the observation that $P(1,t|j,t_0)-P_{\infty}(1,t)$ (for $t\geq t_0$)
fulfills the homogeneous equation obtained by removing the term
$\gamma_2(t)$ from Eq.~(\ref{MEreduced}), with the initial condition
$P(1,t_0|j,t_0)-P_{\infty}(1,t_0)=\delta_{j,1}-P_{\infty}(1,t_0)$.
Thus, the result for $t\geq t_0$ is
\begin{equation}
\label{condprob}
P(1,t|j,t_0)=P_{\infty}(1,t)+\left[\delta_{j,1}-P_{\infty}(1,t_0)\right]
e^{-\Gamma (t-t_0)},
\end{equation}
and, likewise, $P(2,t|j,t_0)=1-P(1,t|j,t_0)$. In terms of the
time-periodic one-time probability in Eq.~(\ref{persol}) the two-time
joint probability reads
\begin{eqnarray}
\label{twotime}
P_{\infty}(j,t;k,t_0)&=&(-1)^{j+k}P_{\infty}(1,t_0)P_{\infty}(2,t_0)\,
e^{-\Gamma (t-t_0)} \nonumber \\&&+P_{\infty}(j,t)P_{\infty}(k,t_0),
\end{eqnarray}
for $t\geq t_0$. Therefore, the asymptotic two-time correlation
function, $\langle x(t) x(t_0) \rangle_{\infty}^{(TS)}=\sum_{j=1}^2
\sum_{k=1}^2 x_j(t)x_k(t_0) \,P_{\infty}(j,t;k,t_0)$, reads
\begin{eqnarray}
\label{twotimecorr}
\langle x(t) x(t_0) \rangle_{\infty}^{(TS)}&=&\left[\Delta x(0)\right]^2
P_{\infty}(1,t_0)P_{\infty}(2,t_0)\, e^{-\Gamma (t-t_0)}\nonumber \\
&&+\langle x(t)\rangle_{\infty}^{(TS)}\langle
x(t_0)\rangle_{\infty}^{(TS)},
\end{eqnarray}
for $t\geq t_0$, where we have used that $\Delta x(t)=\Delta x(0)$.

The two-time correlation function $\langle x(t_0+\tau) x(t_0)
\rangle_{\infty}^{(TS)}$ is a periodic function of $t_0$ with the period
of the external driving \cite {PRA91}. Then, it is convenient to apply a
time-average to obtain the time-homogenous correlation function,
$C^{(TS)}(\tau)$, i.e.
\begin{equation}
\label{Ctau}
C^{(TS)}(\tau)= \frac{1}{T} \int_{0}^{T} \ud t_0 \, \langle x(t_0+\tau)
x(t_0)\rangle_{\infty}^{(TS)}.
\end{equation}

In virtue of Eq.~(\ref{twotimecorr}), $C^{(TS)}(\tau)$ can be written as
the sum of two contributions: a coherent part, $C_{coh}^{(TS)}(\tau)$,
which is periodic in $\tau$ with period $T$, and an incoherent part,
$C_{incoh}^{(TS)}(\tau)$, which decays to $0$ as $\tau \rightarrow
\infty $. These two contributions are given by, respectively,
\begin{equation}
\label{Ccohtau}
C_{coh}^{(TS)}(\tau)= \frac{1}{T} \int_{0}^{T} \ud t_0 \, \langle
x(t_0+\tau)\rangle_{\infty}^{(TS)}\langle x(t_0)\rangle_{\infty}^{(TS)},
\end{equation}
and
\begin{equation}
\label{Cincohtau}
C_{incoh}^{(TS)}(\tau)= \frac{\left[\Delta x(0)\right]^2 e^{-\Gamma
\tau} }{T} \int_{0}^{T} \ud t_0 \, P_{\infty}(1,t_0)P_{\infty}(2,t_0).
\end{equation}

Upon combining Eqs.~(\ref{persol}), (\ref{TSaverag}), (\ref{Ccohtau})
and (\ref{Cincohtau}), one obtains after some cumbersome algebra
\begin{eqnarray}
\label{TSccoh}
C_{coh}^{(TS)}(\tau)&=& (-1)^{n(t)} \Bigg\{\langle x(0)\rangle_{eq}^2
\left[2 n(\tau)+1-\frac{4 \tau}{T} \right] \nonumber\\ &&+\,\frac{4
\,\sech\left(\frac{\Gamma T}{4}\right)\,\sinh \left\{\frac{\Gamma
T}{4}\left[2 n(\tau)+1-\frac{4 \tau}{T} \right] \right\}}{\Gamma
T}\nonumber\\ &&\times \,\left[\left[\frac{x_M(0)}{2} \right]^2-\langle
x(0)\rangle_{eq}^2 \right] \Bigg\},
\end{eqnarray}
and
\begin{eqnarray}
\label{TScincoh1}
C_{incoh}^{(TS)}(\tau)&=&\frac{\left[\Delta x(0)\right]^2 e^{-\Gamma
\tau}}{4} \, \Bigg\{1+\left[\Delta P_{eq}(0)\right]^2\,\nonumber \\
&&\times \,\left[\frac{4 \tanh\left(\frac{\Gamma T}{4}\right)} {\Gamma
T}-1\right]\Bigg\}.
\end{eqnarray}
According to McNamara and Wiesenfeld \cite{McNWie89}, the output SNR,
$R_{out}^{(TS)}$, is defined in terms of the Fourier transform of the
coherent and incoherent parts of $C^{(TS)}(\tau)$ as
\begin{equation}
\label{snr}
R_{out}^{(TS)} =\frac {\lim_{\epsilon \rightarrow 0+}
\int_{\Omega-\epsilon}^{\Omega+\epsilon} \ud\omega\;
\tilde{C}^{(TS)}(\omega)}{\tilde{C}_{incoh}^{(TS)}(\Omega)},
\end{equation}
where $\Omega=2 \pi/T$ is the angular frequency of the external driving,
and $\tilde{H}(\omega)$ denotes the Fourier cosine transform of
$H(\tau)$, i.e., $\tilde{H}(\omega)=2/\pi \int_0^\infty \ud\tau\,H(\tau)
\cos (\omega \tau)$. Note that this definition of the output SNR differs
by a factor $2$, stemming from the same contribution at $\omega = -
\Omega$, from the definitions used in earlier works \cite
{RMP,PRA91}. The periodicity of the coherent part gives rise to delta
peaks in the spectrum. Thus, the only contribution to the numerator in
Eq.\ (\ref{snr}) stems from the coherent part of the correlation
function. The output SNR can then be expressed as
\begin{equation}
\label{snr1}
R_{out}^{(TS)}=\frac{Q_u^{(TS)}}{Q_l^{(TS)}},
\end{equation}
where
\begin{equation}
\label{num}
Q_u^{(TS)}= {\frac 2T} \int_0^T \ud \tau \,C_{coh}^{(TS)}(\tau) \,\cos
(\Omega \tau),
\end{equation}
and
\begin{equation}
\label{den} Q_l^{(TS)}=\frac 2\pi \int_0^\infty \ud \tau
\,C_{incoh}^{(TS)}(\tau) \,\cos (\Omega \tau ).
\end{equation}
Then, from Eqs.~(\ref{num}), (\ref{den}), (\ref{TSccoh}) and
(\ref{TScincoh1}), we get after some simplifications
\begin{equation}
\label{TSnum}
Q_u^{(TS)} =\frac{2\left[4 \langle x(0) \rangle_{eq}^2\Gamma^2
+\left[x_M(0)\right]^2 \Omega^2\right]}{\pi^2 (\Gamma^2+\Omega^2)},
\end{equation}
and
\begin{eqnarray}
\label{TSden}
Q_l^{(TS)} &=&\frac{\left[\Delta x(0)\right]^2 \Gamma}{2 \pi
\left(\Gamma^2+ \Omega^2 \right)}\, \Bigg\{1+\left[\Delta
P_{eq}(0)\right]^2\,\nonumber \\ &&\times \,\left[\frac{4
\tanh\left(\frac{\Gamma T}{4}\right)} {\Gamma T}-1\right]\Bigg\}.
\end{eqnarray}
The signal-to-noise ratio of the input signal, $F(t)+\xi(t)$, can
readily be evaluated from the definition, yielding
\begin{equation}
\label{snrinp}
R_{inp}=\frac{4 A^2}{\pi D}.
\end{equation}
Thus, the SR-gain which is defined as the ratio of the SNR of the output
over the SNR of the input, emerges as
\begin{equation}
\label{gain}
G^{(TS)}=\frac {R_{out}^{(TS)}}{R_{inp}},
\end{equation}
and it can be evaluated explicitly upon combining the set of
Eqs.~(\ref{snr1}), (\ref{TSnum}), (\ref{TSden}), and (\ref{snrinp}).

In Fig.~1 we compare our analytical results for the behavior of several
SR-quantifiers as a function of the noise strength $D$ with results
obtained by numerically integrating the Langevin equation
[Eq.~(\ref{langev})]. The numerical precise solution is based on an
algorithm due to Greenside and Helfand; for details see in the Appendix
of Ref.~\cite{casgom03}. The value of the subthreshold input amplitude
is chosen as $A=0.25$, and the frequency of the external driving is set
at $\Omega=0.01$. This small value of the angular frequency has been
chosen in order to observe a characteristic non-monotonic behavior
versus $D$ of several quantifiers associated to SR and, simultaneously,
SR-gains exceeding unity over a wide range of $D$ (see
Ref. \cite{casgom03new}). The agreement between the analytical nonlinear
response results and the numerical results is surprisingly good at
moderate noise values, and becomes even excellent for small values of
$D$. The better agreement at low noise values corroborates with the fact
that with increasing noise strength the Markovian two-state description
also worsens \cite {Talkner03}.

\begin{figure}
\includegraphics[width=8cm]{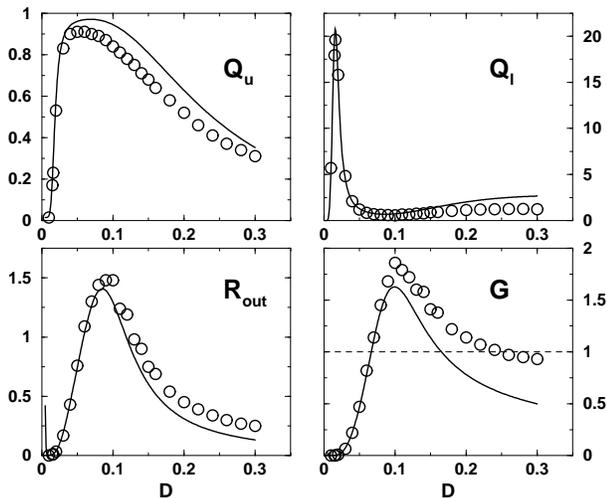}
\caption{\label{wp3} Stochastic Resonance beyond linear response: Several
SR-quantifiers are depicted {\it vs.} the noise strength $D$; namely, the
cosine transform of the coherent part of the driven correlation function
at angular driving frequency $\Omega$, i.e. numerator of the SNR ($Q_u$),
and the cosine transform of the incoherent part, respectively, i.e. the
denominator ($Q_l$), the output SNR ($R_{out}$) and the SR-gain ($G$).
These characteristic quantities are evaluated for a rectangular driving
force with angular frequency $\Omega=0.01$ and subthreshold amplitude
$A=0.25$. The solid lines depict the results obtained within the two-state
description, whereas the numerical precise results obtained from the
driven Langevin dynamics in (\ref {langev},\ref {potential}) are given by
the circles.}
\end{figure}

With this work we have put forward an analytical two-state description for
both, the nontrivial, nonlinear population dynamics and the nonstationary
correlation behavior of  noisy bistable systems driven by periodic
rectangular subthreshold forces. It is indeed remarkable that our
analytical, nonlinear two-state approach does capture well, both the
non-monotonic, bell-shaped behavior of the nonlinear SNR {\it vs.} noise
strength $D$, i.e. the characteristic SR phenomenon, and the occurrence of
SR-gains larger than unity. This latter result is a true benchmark of the
nonlinear response behavior of a driven bistable stochastic dynamics. Our
analytical findings corroborate those obtained by means of numerical
solutions of the Langevin equation \cite{casgom03new}.

This very two-state theory beyond linear response of a driven, metastable
stochastic population dynamics likely proves useful also for phenomena
other than Stochastic Resonance: it equally well can be applied and
generalized to describe the behavior of rocked Brownian motors \cite{PT},
the description of the switching dynamics over adiabatically sloshing
potential landscapes \cite{surmounting} or also to driven noisy threshold
characteristics in general, such as e.g. for a driven neuronal noisy
spiking dynamics.

\begin{acknowledgments}
The authors acknowledge the support of the Direcci\'on General de
Ense\~nanza Superior of Spain (BFM2002-03822), the Junta de
Andaluc\'{\i}a, DAAD-Acciones Integradas (P.H., M.M.) and the
Sonderforschungsbereich 486 of the DFG.
\end{acknowledgments}


\end{document}